\documentclass{ijuc}

\usepackage[utf8]{inputenc}
\usepackage[pdftex]{graphics}

\usepackage{amsmath,amssymb}
\graphicspath{ {./images/} }
\usepackage{caption}
\begin{document}

\title{Simulated Autopoiesis in Liquid Automata}
\author{Steve Battle\email{steve.battle@uwe.ac.uk}}

\institute{Deptartment of Computer Science and Creative Technologies, \\
University of the West of England, Bristol, UK
}

\maketitle
\begin{abstract}
We present a novel form of Liquid Automata, using this to simulate autopoiesis, whereby living machines self-organise in the physical realm. This simulation is based on an earlier Cellular Automaton described by Francisco Varela. The basis of Liquid Automata is a particle simulation with additional rules about how particles are transformed on collision with other particles. Unlike cellular automata, there is no fixed grid or time-step, only particles moving about and colliding with each other in a continuous space/time.
\end{abstract}


\section{Introduction}

Living systems cannot be understood separately from they environment they live in. They are organised in a causal circular process of becoming, and it is this very circularity that is a necessity for living – autopoietic - systems. An autopoietic system is alive if it produces itself in the physical space, based on interactions between physical elements that go on to produce new physical elements necessary for the regeneration of the system \cite{maturana1980}. A living system is a self-referential domain of interactions in the physical space, generally a network of 'chemical' relationships. According to Maturana and Varela, ``An autopoietic machine is a machine organized (defined as a unity) as a network of processes of production (transformation and destruction) of components.'' \cite[p78]{maturana1980}

However, there are many kinds of chemical networks that aren't alive, consider a chemical explosion which exhibits a runaway chain reaction of positive feedback. The signature of life is the emergence of a structure that distinguishes self from non-self, closing it off from its environment, ``A universe comes into being when a space is severed into two. A unity is defined.'' \cite[p73]{maturana1980} This closure emerges from, and is dynamically and homeostatically maintained by, the underlying organisation. There are thus two key components to autopoiesis:

\begin{enumerate}
\item Organisational closure: A network of the processes of production, e.g. a chemical reaction system
\item Structural closure: The appearance of a homeostatically maintained boundary that divides self from non-self, e.g. a cell-wall. This is also described as the maintenance of identity.
\end{enumerate}

\begin{figure}[h]
\begin{center}
\scalebox{0.5}{\includegraphics{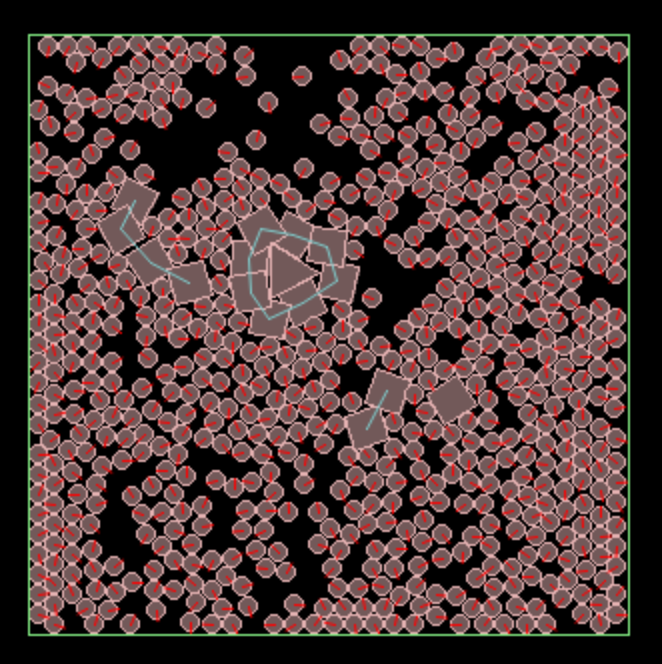}}
\end{center}
\caption{Liquid automaton showing a boundary (blue links) forming around the catalyst (triangle), distinguishing self from non-self. The catalyst transforms the substrate (circles) into its structural building blocks (squares).}
\label{automaton}
\end{figure}

Simulated autopoiesis is a demonstration of how autopoiesis works and enables us to study simpler autopoietic systems not found in nature. We define an organisationally closed system, with the aim of generating structural closure. The system explored here is defined as a organisationally closed chemical reaction system (CRS), bathed in a liquid substrate that provides the raw materials from which the system builds itself. In figure \ref{automaton}, the seed is a catalytic agent (triangle) that transforms the substrate (circles) into its structural building blocks (squares). The system, in effect, feeds off the substrate particles. The link particles are able to self-organise (blue links) themselves into a structure akin to a long-chain polymer. This long chain of bonded links is able to wrap around and close in on itself to form a closed boundary, analogous to a cell wall enclosing the catalyst. 

The basis of this \emph{Liquid Automaton} is a 2D particle simulation with additional rules about how particle state changes on contact with other particles. Unlike cellular automata, there is no fixed grid, just particles moving about and colliding with each other in a continuous space. By analogy with cellular automata, the Liquid Automaton is a variety of \emph{collision-based} system \cite{toffoli2002}. The simulation is based on just three rules, defined below, based on an earlier \emph{cellular} automaton devised by Varela \cite{varela1979} (see section \ref{history}). These rules are invoked on contact (as with rules 1 and 2), or may occur spontaneously (as with rule 3).

\section{Definitions}

A Liquid Automaton, L, is defined as follows. The nomenclature is based on tools for building particle systems including Box2D and LiquidFun.

\begin{equation}
    L = (W,P,B,J,R)
\end{equation}
\begin{align*}
\text{where} &: \\
& W \text{ is a world} \\
& P \text{ is a set of particle Types} \\
& B \text{ is a set of bodies, instances of particles}  \\
& J \text{ is a set of joints, that enable bodies to be bonded together} \\
& R \text{ is a set of reaction rules}
\end{align*}

The world, $W$, defines the space, or coordinate system, in which the liquid automaton operates. There is currently a great deal of interest in 2D particle simulations, and this work builds on those advances, though the same ideas extend to three (and higher?) dimensions. A world may, or may not, have gravity; this simulation does not. The simulation explored here introduces energy into the system in the form of a random `Brownian' motion defined in terms of a Wiener process \cite{nelson67} along the x,y dimensions. The force applied along each dimension is a normally distributed random variable with zero mean, and variance, $ (delta)^2 dt$, correlated with a single parameter delta, and time period, $dt$, which varies dynamically.

The set of particle types, $P$, defines which kinds of particle exist in the liquid automaton. In this example, $P = \{K,S,L\}$, representing the catalyst, the substrate, and the link particle. The reaction rules will be expressed in terms of these particle types. Each particle type has a distinct shape, which may be a compound shape including specific sensor sites or regions that can create a `neighbourhood' around a particle extending beyond its visible outline.

Bodies, $B$, represent the `physical' configuration of the system. Each body $b \in B$ has an associate particle type $p \in P$. Bodies provide the underlying particle simulation having both a position, orientation and velocity in space, as well as other properties including mass. 

Joints, $J$, enable particles to bond together, forming compound particles. This mechanism is analogous to covalent bonding, enabling the formation of `molecular' structures. The underlying particle simulation will typically offer a range of joint types.

The reaction rules, $R$, define the organisation of the system. These rules operate asynchronously, either on collision of two or more particles, or may occur spontaneously at a given rate. Reaction rules should conserve mass. The set of reaction rules that define the simulation of autopoiesis are as follows:

\begin{align}
\text{composition} &: K + 2S \rightarrow K + L \label{eq:composition}\\
\text{concatenation} &: L^{n} + L \rightarrow L^{n+1} \label{eq:concatenation}\\
\text{disintegration} &: L \rightarrow 2S \label{eq:disintegration}
\end{align}
\begin{align*}
\text{where} &: \\
& K - \text{catalyst (triangle)} \\
& S - \text{substrate (circle)} \\
& L - \text{link (square)}
\end{align*}

A reaction rule has a left-hand side defining the \emph{reactants}, separated by an arrow from the reaction \emph{products} on the right-hand side. The appearance of a $+$ (plus) symbol between reactants indicates a collision event where all of the indicated particles have come into contact. Each particle type may be prefixed by a number indicating an integer number of particles of the same type, so that $K + 2S$ is equivalent to $K + S + S$, the interaction of three particles. The use of plus between reaction products, indicates that the reaction produces many outputs. Bonds are indicated by multiplicative operators, such as the dot operator (not shown). Consequently, the use of superscripts, such as $L^{n}$ in rule 2, denotes $n$ particles of the same type, $L$, bonded together (by a joint) \cite{Hordijk2015}. 

In a liquid automaton we can only define explicit reaction rates for spontaneous reactions that don't involve collisions, such as the disintegration rule. The remaining rules are triggered as a result of collisions between different particles. Composition only occurs when a catalyst comes into simultaneous contact with a pair of substrate particles, which are then fused together to form a link particle. Concatenation enables the self-organisation of the boundary. Links are able to make up to two connections, and when two links collide, and are able to do so, they form a bond between them (shown as a blue line). As these links are being formed around the catalyst, it's likely that the bonded links will close the circle, enclosing the catalyst. The embodiment of the catalyst as a shape with a certain extent, creates an internal pressure that holds the links apart, (mostly) preventing them forming useless tight enclosures. These `zombie' structures without a catalyst at their heart, lack the organisation necessary to repair themselves. Links are also subject to decay, and may spontaneously disintegrate back into a pair of substrate particles. Disintegration triggers the homeostatic repair of incomplete boundaries, and also serves to recycle `waste' links that may have drifted too far from the catalyst, along with the aforementioned Zombies. In this experiment, a single catalyst particle is introduced at the beginning, which is not subject to disintegration. Autopoiesis is intriguing because reproduction is not seen as an essential quality of life. Hence, this minimal model is not concerned with the production of new catalysts, or autocatalysis.

\section{History} \label{history}

\begin{figure}
\begin{center}
\scalebox{0.5}{\includegraphics{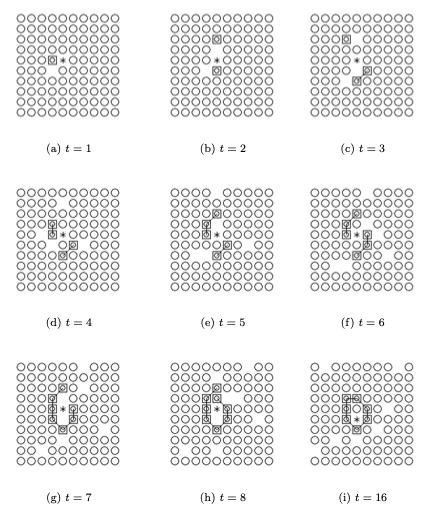}}
\end{center}
\caption{A simulation of autopoiesis using a discrete time cellular automaton on a rectangular grid, based on Varela's original algorithm. In (a) a pair of substrate (circles) are transformed into a single link (squared circle) by the catalyst (asterisk). By (c) we see the first bonds forming, then in (i) these finally form a closed boundary around the 
catalyst. }
\label{fig:ca}
\end{figure}

The work in this paper was inspired by the Cellular Automaton, or tesselation model, described by Francisco Varela \cite{varela1979} as a way to explain the process of autopoiesis in concrete terms. It was intended as a simplified model of the reaction rules found in living cells. The output shown in Figure \ref{fig:ca} is produced by code based closely on an algorithm provided by Varela \cite{Varela74}. While the figures in Varela's paper are hand-drawn based on printed output from a FORTRAN program \cite{McMullin97}, the same graphics are reproduced by our code. The catalyst is shown as an asterisk, substrate particles as circles, and link particles as squared circles. Unusually for a cellular automaton, bonds are indicated by lines drawn between pairs of cells representing bonded link particles.

It implements the same reaction rules as our liquid automaton, as defined in equations \eqref{eq:composition} to \eqref{eq:disintegration}.
Figure \ref{fig:ca} illustrates a number of steps from a single run. In step (a) at time, $t=1$, a pair of substrate particles are transformed into a single link particle by the catalyst. After the production of a number of link particles, we see the first bonds forming between them by step (c) at $t=3$. The composition rules in Varela's algorithm are constrained to forming only obtuse bond angles. This prevents the uppermost link particle at $t=8$ (h) from bonding with the particle immediately below it, so it is only when this particle disintegrates later at $t=16$ (i) that enables the remaining links to re-bond, and form a closed boundary around the catalyst.

These same rules were re-implemented in a later program called SCL (Substrate, Catalyst and Link) by McMullin using the SWARM system \cite{McMullin97b} where chemical reaction rules were captured in a modular fashion enabling their reaction rates to be more precisely controlled. This system also introduced a more configurable way to control the random motion of particles in the 2D space. However, unlike the proposed liquid automata model, the SWARM system modelled space as a square lattice with a toroidal toplogy, and discrete time.

Another cellular automaton, Conway's ``Game of Life'' \cite{Gardner1970,Berlekamp1982,Adamatzky2010,Izhikevich2015} is also amenable to analysis using autopoietic theory \cite{beer2015}. The Game of Life typifies emergent, self-organized behaviour and in these behaviours it is possible to identify processes, local transformations in the space of patterns, that are organisationally closed. Thus we see cyclic transformations of patterns that are self-reproducing, like the classic `blinker' and `glider'. However, the boundary of a self-reproducing form has to be understood in a more abstract way than a cell-wall, as the set of surrounding cells that need to be specifically on or off, for it to survive; its contingent neighbourhood. However, this abstract conception of boundary has no power to contain the form and protect it from external disruption.

\section{Particle Systems}

Particle systems are game physics engines designed to reproduce naturalistic phenomena based on objects moving around, typically in a 2D space. The system used for this simulation described in this paper is Box2D (specifically pybox2D), a rigid body simulation library for games. Interestingly, Box2D has been used as the game engine for a number of implementations of ``Angry Birds.'' Each particle is a 2D body with mass and velocity, so particles have three degrees of freedom; translation in x,y coordinates, and rotation. Each body is associated with one or more shapes which can be any geometrical construct, such as the squares, circles, and triangles used here. Joints define constraints on the relative motion between two bodies, used here to create bonds between neighbouring links. Forces, torques, and impulses are applied to bodies to make them move. Box2D includes a high performance iterative constraint solver that resolves joint constraints, particle motion and resulting collisions \cite{Catto2005}. Particles can bounce off each other in elastic collisions, or slide against each other based on a realistic simulation of the frictional forces between them.

\section{Analysis}

\begin{figure}
\begin{center}
\scalebox{0.5}{\includegraphics{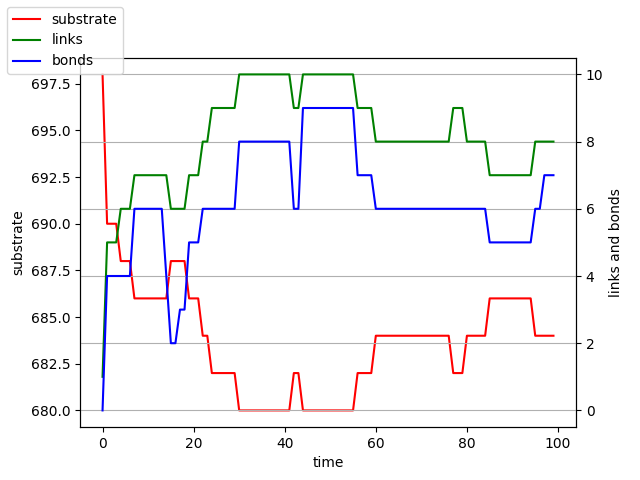}}
\end{center}
\caption{Absolute quantities of substrate, links, and bonds. The quantity of substrate drops rapidly at the start as they collide with the exposed catalyst. Conversely, the number of links increases as they are produced in the reaction of the substrate with the catalyst. Bonds form as links are produced.}
\label{auto-plot}
\end{figure}

For liquid automata to be more than a pretty picture, we must be able to perform some analysis on the system. We must ask ourselves, ``what does the system do?'' Firstly, we can inspect the raw quantities of different elements in the system over time. The system starts out with a single catalyst and a predefined number of substrate particles, which is initialised to 700 as a default. There are initially, no link particles or bonds. Figure \ref{auto-plot} shows the typical progress of the system over the first 100 seconds. The number of substrate particles drops rapidly at the start as they come into contact with the exposed catalyst. Conversely, the number of link particles increases as they are produced in the reaction of the substrate with the catalyst. After a while we begin to see the number of substrate particles increase again, as the effect of the disintegration rule becomes more prominent. As expected, these quantities are perfectly anti-correlated (Pearson correlation coefficient = -1).

As the link particles come into contact with each other they form mutual bonds. There can never be more bonds than link particles, and there will be fewer bonds than links on the whole, except in the unlikely state where the link particles form a perfect ring. We expect these quantities to be correlated; the Pearson correlation coefficient between the links and bonds is 0.68 (with a p-value significant at 1\% level).

\begin{figure}
\begin{center}
\scalebox{0.5}{\includegraphics{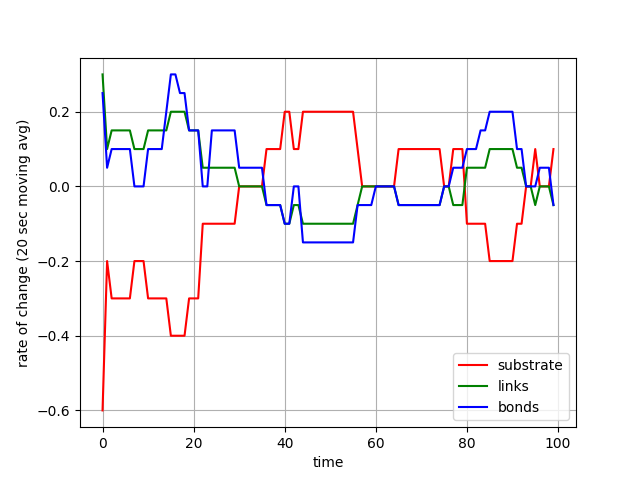}}
\end{center}
\caption{Rates of change of substrate, links, and bonds based on a 20 second moving average. The relationship between rates of change for substrate, $\Delta S$, and links, $\Delta L$, is $\Delta S = -2\Delta L$.}
\label{fig:auto-rates}
\end{figure}

A more informative plot shows the rates of change of each of these variables. Figure \ref{fig:auto-rates} plots the rates of change over the first 100 seconds, using a 20 second moving average; the plot is too spiky to make sense of without averaging. It's easier to observe that for rates of change for substrate, $\Delta S$, and links, $\Delta L$, the relationship between them is $\Delta S = -2\Delta L$, in accordance with the 2:1 ratio of substrate to link composition and disintegration.

These tests merely show that the code is working properly as an implementation of the reaction rules, in order to achieve organisational closure. To evaluate the system's capacity for \emph{self-preservation}, we must look deeper into what the system is actually doing. We hypothesize that the system is acting as a simple feedback controlled regulator, acting to limit the flow of substrate particles to the catalyst. 

\begin{figure}
\begin{center}
\scalebox{0.5}{\includegraphics{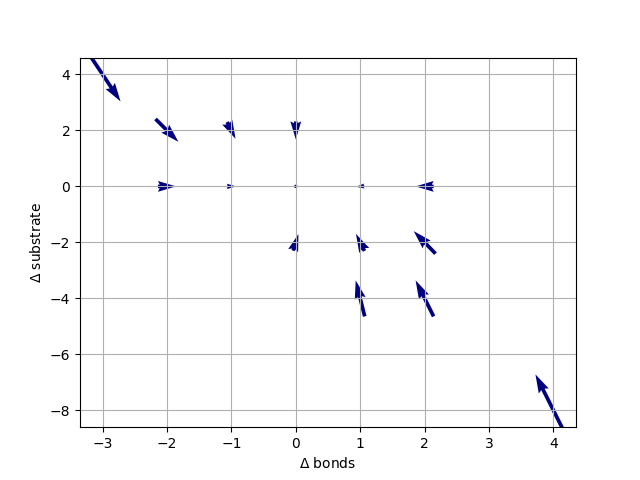}}
\end{center}
\caption{Phase plot of the average rate of change in the number of links versus bonds. The origin is an attractor, with larger vectors at greater distances from the origin. }
\label{fig:auto-phase}
\end{figure}

One way to visualise this is to plot a phase diagram of the system as seen in Figure \ref{fig:auto-phase}. This shows the mean direction of travel at each state visited by the system, in terms of the rates of change of the substrate versus bonds. The rate of change of bonds was selected as the horizontal axis, because a comparable phase plot of substrate against links would lie on a straight line, as there is a simple linear relationship between them. We can think of this diagram as showing the flux of substrate being composed into links, forming bonds, before disintegrating back into substrate. The data points themselves are not averaged, so we see only integer differences in quantity. Observe that the vectors point towards the origin, with larger magnitudes at greater distances from the origin. With more bonds, fusion of substrate particles to form links is throttled back, reducing bond formation. As links disintegrate back into substrate, bonds break, and the flow of substrate to the catalyst increases. The origin is an attractor, and the system therefore tends to move towards dynamic equilibrium around zero substrate flux.

\section{Conclusion}

We presented a Minimal Autopoietic Simulation, demonstrating both organisational and structural closure. This simulation was created within a novel Liquid Automata framework that enables us to define the system in terms of moving bodies within a continuous space/time (of course, realised discretely at some lower level), rather than a fixed grid and discrete time-steps, as in a cellular automaton. It invites investigation of more interesting chemical networks with more sophisticated metabolisms.

\section{Acknowledgments}

I would like to acknowledge the help of my friend and colleague, Dan Brickley, who pulled together the invisible threads that make this work. 

\raggedright
\bibliography{main}
\end{document}